\documentclass{PNASsupp}

\usepackage[dvips]{graphicx}

\usepackage{amssymb,amsfonts,amsmath}
\usepackage[thinspace,thinqspace,squaren]{SIunits}					

\newcommand{\YRS}{YbRh$_2$Si$_2$}
\newcommand{\TN}{\ensuremath{T_{\text N}}}						
\newcommand{\RH}{\ensuremath{R_{\text H}}}						
\newcommand{\TRH}{\ensuremath{\tilde R_{\text H}}}				
\newcommand{\TRHO}{\ensuremath{\tilde R_{\text H}^{0}}}		
\newcommand{\TRHI}{\ensuremath{\tilde R_{\text H}^{\infty}}}
\newcommand{\TRHB}{\ensuremath{\TRH(B_1)}}						
\newcommand{\RHO}{\ensuremath{\RH ^0}}								
\newcommand{\RHI}{\ensuremath{\RH ^{\infty}}}					
\newcommand{\RHB}{\ensuremath{\RH (B_2)}}							
\newcommand{\RT}{\ensuremath{\RH (T)}}								
\newcommand{\rhoH}{\ensuremath{\rho_{\text H}}}					
\newcommand{\drhoHdB}{\ensuremath{\partial \rhoH / \partial B_1}}	
\newcommand{\rhoO}{\ensuremath{\rho ^0}}							
\newcommand{\rhoI}{\ensuremath{\rho ^{\infty}}}					

\contributor{Submitted to Proceedings
of the National Academy of Sciences of the United States of America}
\url{www.pnas.org/cgi/doi/xxx}
\copyrightyear{xxxx}
\issuedate{Issue Date}
\volume{Volume}
\issuenumber{Issue Number}

\begin{document}



\title{Fermi-surface collapse and dynamical scaling near a quantum critical point}





\author{Sven Friedemann\affil{1}{Max Planck Institute for Chemical Physics of Solids, N{\"o}thnitzer Str.~40, 01187 Dresden, Germany}, 
Niels Oeschler\affil{1}{},
Steffen Wirth\affil{1}{},
Cornelius Krellner\affil{1}{},
Christoph Geibel\affil{1}{},
Frank Steglich\affil{1}{},
Silke Paschen\affil{2}{Institute of Solid State Physics, TU Vienna, Wiedner Hauptstr. 8-10, 1040~Vienna, Austria},
Stefan Kirchner\affil{3}{Department of Physics and Astronomy, Rice University, Houston,
Texas 77005-1892, USA}\affil{4}{Max Planck Institute for the Physics of Complex Systems,
N{\"o}thnitzer Str.~38, 01187~Dresden, Germany},
\and Qimiao Si\affil{3}{}}

\contributor{Submitted to Proceedings of the National Academy of Sciences
of the United States of America \hfill Classification: Physical Sciences---Physics}

\maketitle

\begin{article}

\begin{abstract}
Quantum criticality arises when a macroscopic phase of matter undergoes a
continuous transformation at zero {temperature}. While the collective fluctuations at
quantum critical points are being increasingly recognized {as playing} an
important role in a wide range of quantum materials, the nature
of the underlying quantum critical excitations  remains poorly understood. Here
we report in-depth measurements of the Hall effect in the heavy-fermion metal
\YRS, a prototypical system for quantum criticality. 
{We isolate} a rapid crossover of the {isothermal} Hall coefficient 
{clearly connected to} the quantum critical point
{from} a smooth background contribution;
{the latter exists away from the QCP
and is only detectable through 
our studies over a wide range of magnetic field.}
{Importantly}, the width of the critical crossover 
is {proportional to} 
temperature, which {violates} the predictions of conventional theory and 
{is} instead {consistent with} an
energy over temperature, 
$\mathbf{{\textit{E}/\textit{T}}}$, scaling of the quantum critical
{single-electron fluctuation spectrum}.
{Our results provide evidence that the quantum-dynamical scaling 
and a critical Kondo breakdown simultaneously operate 
in the same material.}
{Correspondingly, we infer that 
macroscopic scale-invariant fluctuations emerge from the microscopic 
many-body excitations associated with a collapsing Fermi surface.
This insight}
is expected to be relevant to the 
unconventional finite-temperature behavior in a broad range 
of strongly correlated quantum systems.
\end{abstract}





\dropcap{Q}uantum criticality,
epitomizes the richness of quantum effects  in macroscopic
settings \cite{Natphys_08}.  
The traditional description 
is based on the framework of Ginzburg and Landau~\cite{Ma_book}, which focuses
on the notion of an order parameter,  a classical variable. The order parameter
delineates the symmetry breaking of the macroscopic phases, while its
fluctuations at  ever-increasing length and time scales characterize  the
approach towards a second-order quantum phase transition. For metallic
antiferromagnets, this theory appears in the form of a 
spin-density-wave 
{quantum critical point (QCP)} \cite{Hertz_76,Millis_93}. 
Here, the macroscopic
fluctuations of the order parameter are described by a Gaussian theory at the
fixed point, with a vanishing effective coupling among the collective modes  in
the zero-temperature ($T=0$), zero-{energy} (${E}=0$) and infinite-length
limit. Consequently \cite{Sachdev_book}, the collective fluctuations will
violate ${E}/T$-scaling.

By contrast, a{n}
{unconventional} class of quantum criticality, emerging from studies in recent
years \cite{Natphys_08}, incorporates not only the slow fluctuations of the
order parameter, but also some inherent quantum  modes. For heavy-fermion
metals, the new quantum modes  are associated with a critical breakdown of the
Kondo screening effect and the concomitant single-electron Kondo resonance
excitations \cite{Si_01,Coleman_01,Senthil_04}.  These new critical modes can
lead to a critical field theory  that is interacting, instead of Gaussian, and
the collective fluctuations will satisfy 
${E}/T$-scaling
\cite{Schroeder2000,Aronson1995}. The critical Kondo effect itself is manifested
in the nature of microscopic single-electron excitations, with  the Fermi
surface undergoing a {severe} reconstruction at the QCP.

To date, there has been no experiment to determine  
{that} the critical Kondo
destruction {is} the underlying mechanism for the dynamical ${E}/T$-scaling. In
the heavy-fermion quantum  critical material ${\rm CeCu_{{5.9}}Au_{{0.1}}}$, the
magnetic dynamics have been shown to display such a scaling
\cite{Schroeder2000}.
In this material, however, the unconventional QCP appears only  by 
tuning of chemical doping or pressure \cite{Stockert2007};
consequently it has so far not been possible to probe its Fermi surface and related
single-electron properties with sufficient resolution.
The heavy-fermion system ${\rm YbRh_2Si_2}$ features an unconventional QCP that
is  accessible by {the} application of a relatively small magnetic field
\cite{Trovarelli_00,Gegenwart2002}, thereby allowing the study {of magnetotransport}
across the QCP.
While indications of a rapid Fermi surface change  in ${\rm YbRh_2Si_2}$ has
appeared through the observation of  a crossover in the Hall effect
\cite{Paschen2004}, no information has been extracted on the dynamical
fluctuation spectra of either the magnetic or single-electron excitations.
Moreover, the Hall crossover has alternatively been interpreted in terms of a
background contribution in the non-magnetic heavy-fermion phase  through
either minute valence variations \cite{Norman_05} or Zeeman splitting of
the bands  {\cite{Loehneysen_RMP,Julian_08}}, leaving the nature of the  quantum critical
single-electron excitations uncertain. 
{In addition, the observation of sample dependences in the low-temperature
Hall coefficient raises the important question of how these affect {the Hall crossover}\cite{Friedemann2008}.}
To resolve {these} fundamental issue{s}, {we 
carry out comprehensive, in-depth}
Hall-effect measurements over a wide range of the control parameter, 
the magnetic field, down to {very low} temperatures.
We establish a sample-dependent background component of the 
Hall crossover, {which in turn allows} us to isolate a critical component of the crossover
with properties that are sample-independent.
In addition, we identify a robustly linear-in-temperature 
width of the critical Hall crossover,
{which is compatible with a} {quantum-}{dynamical}
scaling {of} the critical single-electron {excitations}.
Our findings lead to an
unexpectedly direct linkage between the  scale-invariant macroscopic 
fluctuations and the microscopic physics of a collapsing Fermi surface.

\section{Results}
We study the magnetotransport in tetragonal ${\rm YbRh_2Si_2}$ using  a
crossed-field setup, in which two external magnetic fields are applied in
perpendicular directions. This separation
allows for a disentanglement between field-tuning of 
ground states through $B_2$ and {generation of} Hall {response}
through $B_1$:
One field, $B_1$, along the magnetic hard $c$
axis and perpendicular to the electrical current, is used to extract the
initial slope of the Hall resistivity, {$\rho_{\text H}$}, \textit{i.e.}\  the
linear-response Hall coefficient, $R_{\mathrm H}$ 
{(see Supplementary Information I)}. The {second} field, $B_2$,  applied
within the {magnetic{ally} easy} $ab$ plane and  along the {current} direction, 
is used as the control parameter that tunes the system 
{from an antiferromagnetic ground state at low fields 
{across} the QCP towards a high-field paramagnetic state.
The adjacent phases on both sides of the QCP
obey Fermi liquid properties, like a quadratic temperature dependence of the resistivity \cite{Gegenwart2002}.}
{We consider two samples, which span the whole range 
of sample dependences in the Hall coefficient (see Supplementary Information II).}

{Figure~\ref{fig:RHvsB2_new}} shows the  isothermal linear-response Hall
coefficient {of our highest-quality sample} as a function of $B_2$.
 Two features are evident. First, for 
{$B_2$ much larger than the quantum critical field, $B_{2c}$,} the Hall coefficient shows a sizable variation with
the magnetic field; within the experimental error it is linear in $B_2$.
This background feature is likely 
due to Zeeman splitting \cite{Julian_08} {since no indication for a}
valence change \cite{Norman_05} {has been} observed.
The identification of this background
feature is only possible because we have measured {in} a
{substantially} extended range of $B_2$ 
{(see Supplementary Information I)}.
{Second}, there is a sharp crossover feature that rides on top of the background
contribution. This sharp feature is located near $B_{2c}$, and will henceforth be termed the critical Hall-crossover
component.

\begin{figure}  
	\includegraphics[width=.9\columnwidth]{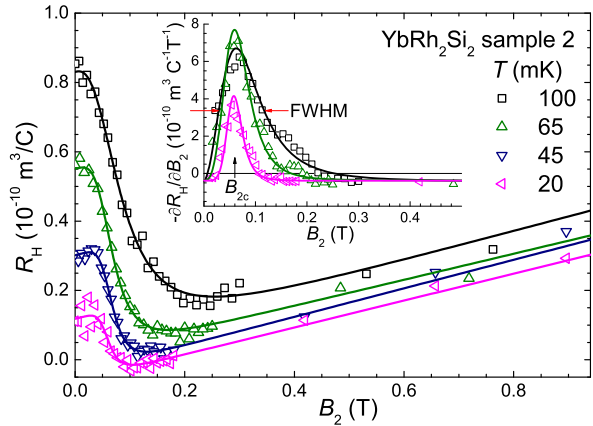}
	\caption{\label{fig:RHvsB2_new} 
	{\bf Crossed-field {Hall-effect} results of \YRS.} Selected
	isotherms of the initial-slope Hall coefficient  $R_{\mathrm H}$ 
	as a function of $B_2$ for sample 2 
	{[which has the smallest residual resistivity (cf. Supplementary Information II)]}.
	The solid lines are best fits
	of  the empirical crossover function {given in eq.~\ref{eq:crossover}} 
	in the Methods section, 
	{extending up to 2\,T}. 
	The anomalous contribution to the Hall effect can be
	neglected as explained in the Supplementary Information III. {The inset illustrates the}
	decomposition of the 
	crossover {in $R_{\mathrm H}(B_2)$} into the critical and background components.
	Here, $-\partial R_{\mathrm H} (B_2)/\partial B_2$ is plotted as a function of $B_2$  together
	with the derivatives of the fitted functions (solid lines). The background
	crossover term corresponds to the  non-zero constant {offset.} 
	The critical crossover term is represented by the sharp
	peak near $B_{2 \text c}$ 
	{(marked by vertical arrow)},  
	whose full-width at half-maximum, FWHM, is defined
	as the crossover width  (specified for one temperature by the {red} horizontal
	arrows).  Standard errors of $R_{\mathrm H}$ are typically of the size of the symbols. }
\end{figure}

{The inset of} Fig.\ \ref{fig:RHvsB2_new} further illustrates the systematic
decomposition of the Hall crossover into the background and critical
components.
{It} plots $-\partial R_{\mathrm H}/ \partial B_2$ as a function of $B_2$. 
The background term appears
as a{n underlying} non-zero {offset},
 while the critical term {manifests itself}
as a sharp peak near $B_{2c}$. More quantitatively,
{Fig.~\ref{fig:RHvsB2_new} shows the separation of the two components
using a fitting procedure specified in the Methods section.}

\begin{figure}       
	\includegraphics[width=.9\columnwidth]{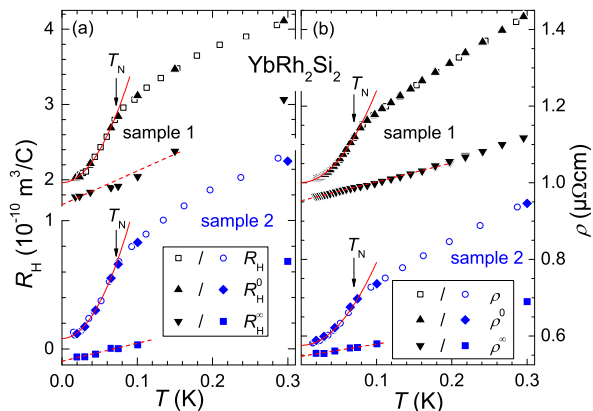}
	\caption{\label{fig:RHvsTlin}
	\textbf{Limiting values of the Hall and magnetoresistivity crossover.}
	\textbf{(a)} Fit parameters $R_{\text H}^0$ and $R_{\text H}^\infty$ of the 
	crossover {in $R_{\mathrm H}$}
	plotted for sample 1 and sample 2
	as a function of temperature together 
	with the {measured} initial-slope Hall coefficient {$R_{\mathrm H}$}. 
	The residual resistance ratios are 70 and 120 
	for sample 1 and sample 2, respectively.
	\textbf{(b)} 
	{Corresponding quantities $\rho^0$ and $\rho^{\infty}$}
	from the analogous analysis
	of the magnetoresistivity crossover 
	(see Supplementary Information IV).
	{Solid lines correspond to fits of a 
	quadratic temperature dependence below $T_{\text N}$ 
	(see Supplementary Information V){,
	as already observed previously {for $\rho(T)$} \cite{Gegenwart2002}}. 
	Dashed lines are guides to the eye.}
	{Arrows indicate the N\'eel temperature.}
	Standard {deviations} are smaller than the 
	symbol size. }
\end{figure}

The critical component is characterized by the difference between $R_{\mathrm H}^0$, the
Hall coefficient before the crossover, and $R_{\mathrm H}^\infty$, the Hall coefficient
after the crossover. {The temperature  dependence of 
$R_{\mathrm H}^0$ and $R_{\mathrm H}^\infty$  in the low-$T$ range
{for both samples} is shown in Fig.~\ref{fig:RHvsTlin}(a)}.
{The experimental finding of a pronounced quadratic temperature dependence of
$R_{\mathrm H}^0$ below $T_{\mathrm N}$ 
allows a proper extrapolation to $T\to 0$ 
yielding a} \emph{finite difference} between 
$R_{\mathrm H}^0$ and $R_{\mathrm H}^\infty$ persisting to zero
temperature {(see Supplementary Information V)}. 
This difference is naturally associated with a change 
of the Fermi surface.
The \emph{magnitudes} of $R_{\mathrm H}^0$ and $R_{\mathrm H}^\infty$, 
{on the other hand, are different for the different samples
rendering the sample dependences of the Hall coefficient 
a common property of {the two} phases at either side of the QCP.}
{Recent {\it ab initio} calculations of the Hall coefficient in
${\rm YbRh_2Si_2}$ suggest that these sample dependences are the effect
of multiple Fermi surface sheets. The "small" (4f-core) and "large" (4f-itinerant)
Fermi surfaces at fields below and above $B_{2\text c}$ in ${\rm YbRh_2Si_2}$ are respectively dominated by two hole and
one hole/one electron Fermi surface sheets \cite{Friedemann2008.2}. Correspondingly,
the step of $R_{\mathrm H}$ as $B_2$ increases through $B_{2\text c}$
is expected to be negative, as is indeed seen here.}

\begin{figure}     
	\includegraphics[width=.9\columnwidth]{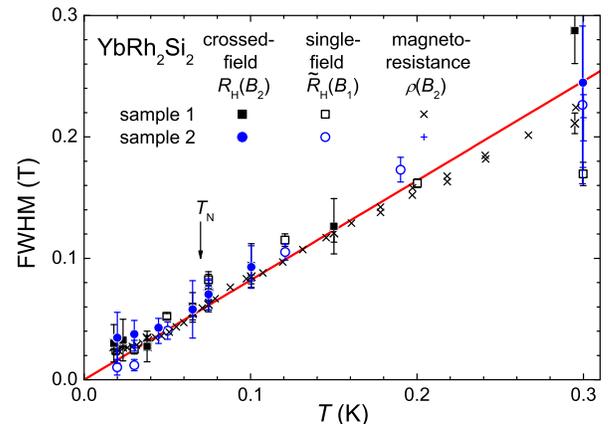}   
	\caption{\label{fig:FWHM_YRS}
	{\bf {Full-width} at half-maximum, FWHM, of the Hall crossover.} The width
	was determined from the derivatives of the fits to $R_{\mathrm H}(B_2)$ in the
	crossed-field setup, to the simultaneously measured magnetoresistivity $\rho(B_2)$,
	and to $\tilde{R}_{\text H}(B_1)$
	of the single-field experiment, respectively.
	{See inset of Fig.~\ref{fig:RHvsB2_new} for the definition of the FWHM.} The values for
	$\tilde{R}_{\text H}(B_1)$ were scaled by $1/11 {=B_{2c}/B_{1c}}$
	(Ref.~\cite{Gegenwart2002})
	to account for the  $c$ axis vs.\ $ab$
	plane magnetic anisotropy of ${\rm YbRh_2Si_2}$.
	The solid line represents a
	linear fit to all data sets
	{with the magnetoresistivity data being well described up to 1\,K}{,
	see Fig.~S6 in the Supplementary Information}.  
	Within the {experimental accuracy} this fit intersects the
	ordinate at the origin. Where there is overlap, our magnetoresistivity results
	are in good agreement with the FWHM directly extracted from the derivative of
	$\rho(B_2)$ presented in Ref.~\cite{Gegenwart2007}. 
	{The crossed-field data obtained earlier in a very limited temperature range \cite{Paschen2004} 
	are in good agreement with both our results and the linear fit. 
	The different temperature dependence found earlier was dominated by the former 
	single-field results differing from ours. This difference is likely a result 
	of an improved orientation procedure that only became possible in a new setup 
	(see Supplementary Information I).}
	{Arrow indicates the N\'eel temperature.} 
	Error bars are standard {deviations}.}
\end{figure}

\begin{figure}        
	\includegraphics[width=.9\columnwidth]{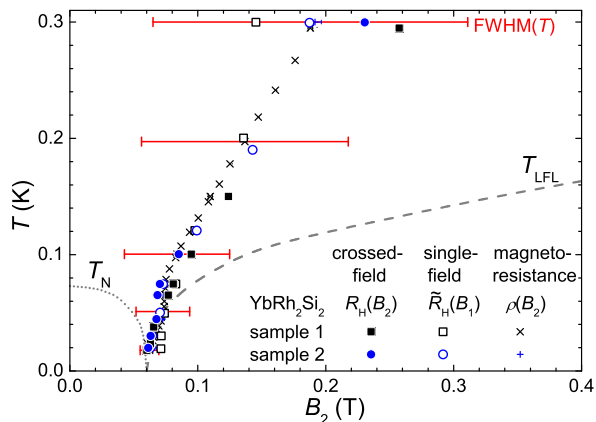}   
	\caption{\label{fig:PD}
	{\bf  Position of the Hall crossover in the {temperature-field} phase
	diagram of \YRS}. The crossover fields, $B_0$, are extracted from
	fits to $R_{\mathrm H}(B_2)$ of the crossed-field experiment,
	to $\tilde{R}_{\text H}(B_1)$ of the single-field experiment,
	{and to $\rho(B_2)$}  {(cf.~Methods)}. The values of the single-field Hall-effect experiment were scaled  by $1/11$ to account for the
	magnetic anisotropy of ${\rm YbRh_2Si_2}$. 
	The red horizontal bars reflect
	the FWHM at {selected} temperatures determined by the fit in
	Fig.~\ref{fig:FWHM_YRS} showing that the crossover fields of the various
	experiments and samples all lie within the  range spanned by the FWHM. The
	dotted (dashed) line represents the boundary of  the antiferromagnetic phase (Fermi liquid
	regime) taken from  Ref.~\cite{Gegenwart2002}.  Error bars are omitted in
	order to avoid confusion with  the width of the crossover; with {the} exception of the
	data  at 0.3\,K and the single-field result of sample 2 at 0.19\,K  the
	{standard} {deviations} are smaller than the symbol size. }
\end{figure}

By contrast, the crossover position and the crossover width  of the critical
component show essentially no sample dependence within the experimental error.
This is seen in Fig.~\ref{fig:FWHM_YRS}, which plots
the full-width {at} 
half-maximum, FWHM, of  {$\partial R_{\mathrm H} / \partial B_2$ isotherms
(Fig.~\ref{fig:RHvsB2_new}(b)), 
and in Fig.~\ref{fig:PD}, which {depicts} the crossover field{, $B_0$,} extracted from  the
fits to $R_{\mathrm H}(B_2)$ for a range of low temperatures in the temperature-magnetic
field phase diagram.}

To corroborate {this fundamental finding}, we have carried out two
additional measurements. The standard single-field Hall-effect setup is used to
{monitor the differential Hall coefficient $\tilde{R}_{\text H}$ 
as a function of the magnetic field $B_1$ applied along the crystallographic $c$ axis (see 
{Supplementary Information I)}.}
{In addition, the} magnetoresistivity, $\rho$, is measured
as a function of a single field, $B_2$, applied within the $ab$ plane. Both
$\rho(B_2)$ and {$\tilde{R}_{\text H}(B_1)$}
can similarly be decomposed into background and
critical terms (see Supplementary Information IV), with the critical {crossover}
terms occurring near the basal-plane critical field, $B_{2\mathrm c}$,
and near the $c$-axis critical field,
$B_{1c}$, respectively;
the ratio $B_{2c}/B_{1c}$ will be used as the  anisotropy ratio 
to convert the
$B_1$ scale into an equivalent $B_2$.  
The {zero-field and high-field values extracted from fits of the crossover function (eq.~\ref{eq:crossover})} to 
magnetoresistivity  {($\rho^0$ and $\rho^\infty$)}
and differential Hall coefficient {($\tilde{R}_{\text H}^0$ and $\tilde{R}_{\text H}^\infty$)}
are presented in Fig.~\ref{fig:RHvsTlin}(b) and in Fig.~S7 
of the Supplementary Information{, respectively}.
{Each quantity} shows a similar sample dependence: 
{As found for the crossed-field results,} {the differences}
$\rho^0-\rho^\infty$ and $\tilde{R}_{\text H}^0-\tilde{R}_{\text H}^\infty$ {remain} finite
in the zero-temperature limit {even though}
 the individual quantities differ for the different samples.
The crossover positions extracted from all the
properties are compiled in Fig.~\ref{fig:PD}. They are largely compatible with
each other, falling within a range spanned by the FWHM;  they define a crossover
energy scale 
{(the $T^*$-line)} \cite{Paschen2004,Gegenwart2007}. 
Finally, the {FWHM of the}
crossover {in 
$\tilde{R}_{\mathrm H}(B_1)$ and $\rho(B_2)$}
{closely follow that of $R_{\mathrm H}(B_2)$ (see Fig.~\ref{fig:FWHM_YRS})}.
We note that the onset of the quadratic form of $R_{\mathrm H}^0(T)$ at $T_{\mathrm N}$ (Fig.\ \ref{fig:RHvsTlin})
is not accompanied by a similar signature in the FWHM at $T_{\mathrm N}$ 
(Fig.~\ref{fig:FWHM_YRS}, see Supplementary Information IV and V).
Therefore, the FWHM 
extrapolates to zero for $T \to 0$ 
implying a jump of all three quantities {($R_{\mathrm H}(B_2)$,
$\tilde{R}_{\mathrm H}(B_1)$ and $\rho(B_2)$)} at the QCP.

The sample independence of both the crossover position and the crossover FWHM
{strongly indicates that}
all the magnetotransport crossovers  manifest the same underlying physics.
Combined with the jump of the Hall coefficient and magnetoresistivity in the
zero-temperature limit, they {imply} the 
interpretation in terms of a sharp Fermi-surface reconstruction at the 
magnetic QCP \cite{Paschen2004}  over that based on the smooth physics of heavy
quasiparticles \cite{Norman_05,Julian_08}. 

Having isolated the critical component of the Hall crossover from  the
background term,  we are now in the position to discuss the detailed nature of
the QCP. For this purpose, we 
have not only carried out crossed-field and single-field Hall and (single-field)
magnetoresistivity measurements over an extended field range for each
temperature, but have also done so for 
{a large set of} temperatures in the low-$T$ range. 
These efforts allow us to reach {the} important and
entirely {new} conclusion 
that the  crossover FWHM is {proportional to}
temperature (Fig.~\ref{fig:FWHM_YRS}).

The Fermi surface is a property of the single-electron excitation
spectrum. In any Fermi liquid, it spans the momenta, ${\bf k}_{\mathrm F}$, at which the
energy dependence of the single-electron Green's function develops a pole at the
Fermi energy. 
{Hence}, a reconstruction of the Fermi surface across the QCP
implies that the single-electron Green's function contains  a singularity
at the QCP.
Indeed, the conduction-electron Green's function  of a Kondo lattice
system can
very generally be written as 
\begin{eqnarray}
G ({\bf k},{E},T) = \frac{1} {{E} -\epsilon_{\bf k} - 
\Sigma ({\bf k},{E},T)}.
\label{G-k-omega}
\end{eqnarray}
In the absence of static Kondo screening, {the self-energy} $\Sigma ({\bf
k},{E},T)$ is non-singular. Correspondingly, the Fermi surface is smoothly
connected to that of the conduction electrons alone; it is
small~\cite{Gegenwart08}. In the
presence of static Kondo screening, $\Sigma ({\bf k},{E})$ develops a pole; for
${E}$ and $T$ small compared to the coherent Kondo scale, it takes the form
\begin{eqnarray}
\Sigma ({\bf k},{E},T) = \frac{(v^*)^2}{{E}-\epsilon_f^*}
+\Delta \Sigma ({\bf k},{E},T).
\label{Sigma-kondo}
\end{eqnarray}
Here $v^*$ and $\epsilon_f^*$ specify the  strength and energy 
of the Kondo resonance \cite{Coleman_review}, and  $\Delta \Sigma ({\bf
k},{E},T)$ is the non-singular term {of the self-energy}. The existence of this pole in $\Sigma
({\bf k},{E},T)$ shifts the Fermi momenta from {their positions on}
the small Fermi surface, ${\bf k}_{\mathrm F}$, to those on a large Fermi surface, ${\bf
k}_{\mathrm F}^*$.
{Approaching} the {point of} critically-destroyed Kondo effect,
the quasiparticle weight vanishes \cite{Si_01, Coleman_01} 
in accordance with the divergence of the quasiparticle 
mass seen in specific heat and resistivity \cite{Gegenwart2002,Gegenwart08}.
In particular, at the QCP, 
both {the strength,} $v^*$, and {the energy,} $\epsilon_f^*$, 
{of the Kondo resonance}
{(see eq.~\ref{Sigma-kondo})}
go to zero in the ${E} \rightarrow 0$ and $T \rightarrow
0$ limit. 
Moreover,
the interacting nature of the fixed point implies an
${E}/T$-scaling 
of the single-electron Green's function:
the reasoning is analogous to that for the
dynamical spin susceptibility \cite{Si_01, Schroeder2000}, 
and the property can be illustrated by explicit calculations in
simplified model settings for critical Kondo destruction \cite{Kirchner_04}.
Similar forms of dynamical scaling of the single-electron spectra are 
likely
a generic feature of other types of Kondo-destroying  QCPs
\cite{Coleman_01,Senthil_04,Pepin}; they appear in related
contexts as well
\cite{MFL89,Anderson_06,Senthil_08}. 
It is worth noting that the linear-in-temperature electrical resistivity
cannot be used as evidence for $E/T$-scaling of the single-electron 
excitations. Indeed, the temperature dependence of the electrical resistivity
does not in general measure the temperature dependence of the single-electron 
relaxation rate and this is so even for 
spin-density-wave
QCPs \cite{Rosch00}.

{The scaling form for the single-electron
Green's function 
in the quantum-critical regime can be expressed as follows:}
\begin{eqnarray}
G({\bf k}_{\mathrm F},{E},T) \sim \frac{1}{T^{\alpha}}
g ( {\bf k}_{\mathrm F}, \frac{{E}}{T}) .
\label{G-scaling}
\end{eqnarray}
The associated relaxation rate, defined in the 
quantum relaxational regime  (${E}  \ll k_B T$)
according to
\begin{eqnarray}
\Gamma({\bf k}_{\mathrm F},T) \equiv \left [-i \partial \ln G ({\bf k}_{\mathrm F},{E},T)
/ \partial {E} \right ]_{{E}=0}^{-1},
\label{Gamma-def}
\end{eqnarray}
is linear in temperature:
$\Gamma({\bf k}_{\mathrm F},T) = c T$, where $c$  is a universal constant. 

\section{Discussion}
{We can use}
these properties of the single-electron Green's function
{to understand the crossover of the Hall 
coefficient. In the Fermi-liquid regimes on either side of the 
QCP, the Hall coefficient reflects the 
respective Fermi surface; it is in particular independent 
of the quasiparticle residue \cite{Khodas_03} (see Supplementary
Information VI). 
The distinct (large and small) Fermi surfaces in the 
two Fermi-liquid regimes yield different end values of the 
Hall coefficient.
The central question is how the two Fermi 
surfaces are connected across the QCP. 
Because the single-electron Green's function characterizes each of the 
two Fermi liquids, this is
related to the critical relaxation
rate, $\Gamma({\bf k}_{\mathrm F},T)$,
of the single-electron states.
At zero temperature, $\Gamma({\bf k}_{\mathrm F},T=0)$ vanishes; the change
from one Fermi surface to the other is sharp, occurring precisely at the QCP.
The Hall coefficient must undergo a sharp jump {in accordance with the experimental findings}. 
{At any non-zero temperature, 
a continuous crossover from one Fermi surface to the other is
controlled by the single-electron relaxation
rate $\Gamma({\bf k}_{\mathrm F},T)$. Given the above described 
behavior of the Hall coefficient
in the adjacent Fermi-liquid
regimes with well defined but different 
Fermi surfaces, its crossover has to be related to the 
finite-temperature broadening of the critical single-electron
states on the Fermi surface. 
{Our observation of a linear-in-temperature
width of the critical Hall crossover is therefore
consistent with a linear-in-temperature relaxation rate.}
By contrast, {our experimental finding}
is incompatible with the
{spin-density-wave} picture of order parameter fluctuations and the concomitant 
Gaussian fixed point, which would be accompanied by a superlinear temperature
dependence of the Hall crossover width (see section VI of the Supplementary
Information)}.

{The} single-electron Green's function serves as the proper means to
specify whether a metal obeys the standard theory of solids -
Landau's Fermi-liquid theory. The fact that eq.~%
(\ref{G-scaling}), with a
non-zero $v^*$, \textit{i.e.}, a large Fermi surface across the QCP, fails
to describe our data is consistent with a breakdown  of the heavy-Fermi-liquid
quasiparticles at the QCP. More generally eq.~%
(\ref{G-scaling}), reminiscent of
the Green's function of gapless interacting electrons in one dimension
\cite{Orgad_01}, invalidates any Fermi-liquid description. By using a single 
set of measurements on the same compound to probe both the collective
fluctuations of the QCP 
and a critical destruction
of the single-electron excitations, our work provides the most direct
association between  quantum criticality and non-Fermi-liquid behavior.

{In summary, we have carried out in-depth magnetotransport measurements in a
prototypical quantum critical heavy-fermion metal, and we are able to
distinguish a robust critical crossover from {a sample-dependent}
background {feature}. By zooming into the  vicinity of the QCP, we have
shown that the width of the critical crossover is not only independent {of
sample quality} but also 
proportional to temperature.}
This proportionality is consistent with the
${E}/T$ form {in the} dynamical critical scaling. Coupled with the fact that
the vanishing width in the zero-temperature limit implies a jump in the Fermi
surface, our findings  point to the microscopic many-body excitations of a
collapsing Fermi surface as underlying the dynamical ${{E}/T}$-scaling  of
the macroscopic critical fluctuations. 
{Our results further establish the $T^{\star}$-line as a means to probe the Kondo breakdown.
{This should hold even when the Kondo breakdown}
is separated from the paramagnetic-to-antiferromagnetic 
QCP \cite{Friedemann2009}.}
In addition, they might help {to} understand why the two coincide in 
stoichiometric \YRS\ and its close vicinity.
More generally, the linkage between microscopics and macroscopics}
is expected to be broadly
relevant to the physics of strong correlations, 
{considering that the finite-temperature properties 
are invariably abnormal in a wide array of quantum materials, and}
given that the Fermi surface and
its evolution as a function of control parameters -- {\it e.g.} from the
underdoped high-temperature cuprate superconductors to the overdoped ones
\cite{Taillefer07,Hussey08} -- are playing an increasingly central role in 
{understanding} {these systems}.



\begin{materials}
The Hall crossovers in both the crossed-field 
{($R_{\mathrm H}(B_2)$)}
and single-field {($\tilde{R}_{\mathrm H}(B_1)$)} 
experiments (see Supplementary Information I)  were fitted with
the empirical crossover function
\begin{eqnarray}
R_{\mathrm H}(B)=R_{\mathrm H}^{\infty}+m B-\frac{R_{\mathrm H}^{\infty} -R_{\mathrm H}^0}{1+(B/B_0)^p}
\label{eq:crossover}
\end{eqnarray}
that contains not only a critical component \cite{Paschen2004}  but also a
linear term $m B$ to account for the background behavior.  
{$R_{\mathrm H}^0$ and $R_{\mathrm H}^{\infty}$ are the zero-field and {infinite}-field values, respectively.}
{The differential Hall coefficient and the magnetoresistivity curves 
were analyzed analogously leading to the corresponding parameters 
$\tilde{R}_{\mathrm H}^0$, $\tilde{R}_{\mathrm H}^{\infty}$ and  $\rho^0$,  $\rho^{\infty}$, respectively.}
{By fitting eq.\ \ref{eq:crossover} to isotherms taken at different temperatures,
the temperature dependences of the parameters were extracted.}
The FWHM {was} extracted from the derivative of the fitted
function as illustrated in {the inset of} Fig.~\ref{fig:RHvsB2_new}.
\end{materials}


\begin{acknowledgments}
We acknowledge discussions with  P.\ Gegenwart,
A.\ Rosch, and A.\ Schofield. Part of the work at Dresden was supported by the
DFG Research Unit 960 ``Quantum Phase Transitions''. {S.~P.\ acknowledges
funding from the European Research Council under the European Community's 
Seventh Framework Programme (FP7/2007-2013)/ERC grant agreement
n$^{\circ}$~227378.} S.~K.\ and Q.~S.\ were supported by the NSF and
the Welch Foundation Grant No. C-1411.
\end{acknowledgments}

\end{article}
\renewcommand{\thefigure}{S\arabic{figure}}
\renewcommand{\theequation}{S\arabic{equation}}
\newpage
\noindent
\textbf{\huge \sffamily Supplementary Information for ``Fermi-surface collapse and dynamical scaling near a quantum critical point''}
\vspace{\baselineskip}

\noindent
\textbf{
Sven Friedemann$^1$, 
Niels Oeschler$^1${},
Steffen Wirth$^1${},
Cornelius Krellner$^1${},
Christoph Geibel$^1${},
Frank Steglich$^1${},
Silke Paschen$^2$,
Stefan Kirchner$^{3,4}$,
\and Qimiao Si$^3$
}
\newline {\small
$^1$ {Max Planck Institute for Chemical Physics of Solids, N{\"o}thnitzer Str.~40, 01187 Dresden, Germany}
$^2$ {Institute of Solid State Physics, TU Vienna, Wiedner Hauptstr. 8-10, 1040~Vienna, Austria}
$^3${Department of Physics and Astronomy, Rice University, Houston, Texas 77005-1892, USA}
$^4$ {Max Planck Institute for the Physics of Complex Systems, N{\"o}thnitzer Str.~38, 01187~Dresden, Germany}
}
\vspace{\baselineskip}

\begin{article}

%
%
\section{Experimental Details}
\label{sec:expDetails}
\begin{figure*}
	\includegraphics[width=.9\textwidth]{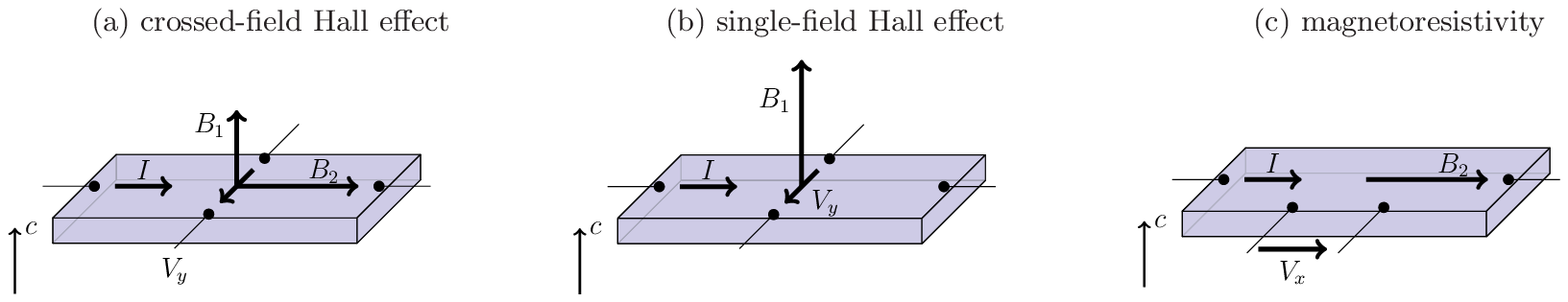}
	\caption{\label{fig:setup} \textbf{Experimental setups.} The experimental setups of the crossed-field and single-field Hall effect and of the magnetoresistivity measurements are depicted (from left to right).}
\end{figure*}

For the low-temperature Hall effect and
magnetoresistivity measurements,  a dilution refrigerator was utilized ($T \geq
\unit{18}\milli\kelvin$). 
A solenoid along with a split-coil magnet allowed
for Hall effect measurements in the crossed-field geometry {which is sketched in Fig.\ \ref{fig:setup}(a)}. {In this setup the dual role of the magnetic field is unraveled by using one field ($B_1$) to generate the Hall response and another field ($B_2$) to tune the ground state of the sample across the quantum critical point.}

\begin{figure}[b]
	\centering
		\includegraphics[width=.7\columnwidth]{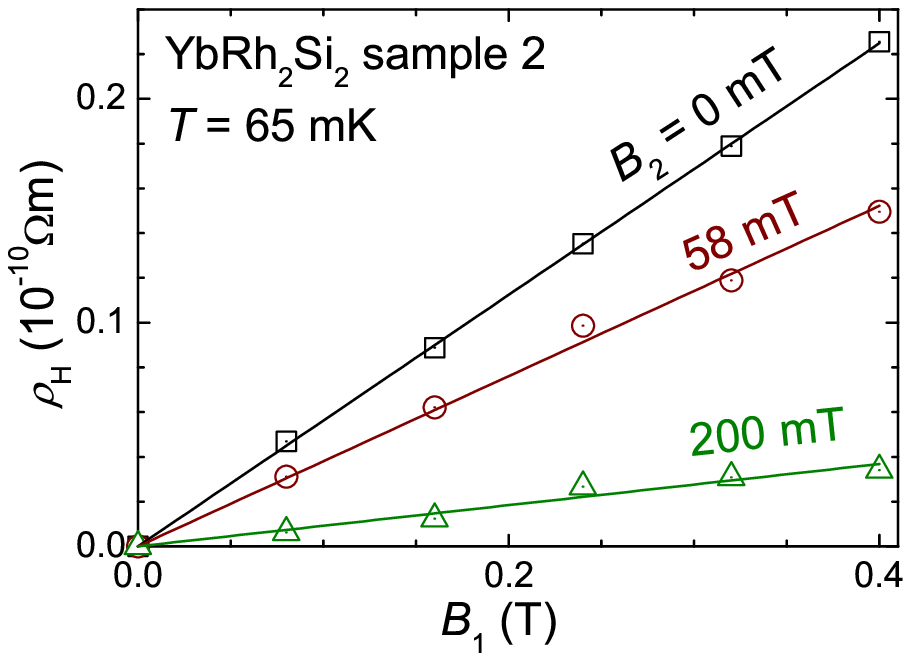}
	\caption{\label{fig:RhoHvsB1_new} \textbf{Calculation of the initial-slope Hall coefficient.} The figure
shows typical isotherms of the Hall resistivity $\rho_{\mathrm H}(B_1,B_2)$ at selected fields
$B_2$. The solid lines herein are linear fits to the data used to calculate the
linear-response Hall coefficient, {$R_{\mathrm H}(B_2) = \rho_{\mathrm H}(B_1,B_2)/B_{1}$. 
}}
\end{figure}

{The small magnetic field ($B_1$) provided by the solenoid is used to produce the linear-response Hall effect. Therefore it was oriented along the magnetic hard $c$ axis and perpendicular to the electrical current flowing within the crystallographic $ab$ plane. Consequently, the Hall voltage $V_y$ is generated transverse to the current within the $ab$ plane. The induced voltages were amplified by low-{temperature} transformers and the signals were measured  by a standard lock-in technique.} We extracted the Hall resistivity {$\rho_{\mathrm H}$}
 as the antisymmetric component of the  field-reversed transversal voltage, $\rho_{\mathrm H}(B_1{,B_2})=t\left[V_y(+B_1{,B_2})-V_y(-B_1{,B_2})\right]/2I$ 
{(with $t$ being the thickness of the sample)}. The linear-response 
Hall coefficient $\RH(B_2)$ was subsequently {derived} from the initial slope of the Hall resistivity $\rhoH(B_1,B_2)|_{B_2}$
\begin{equation}
	R_{\mathrm H}(B_2)\equiv \lim_{B_{1} \rightarrow 0}\rho_{\mathrm H}(B_1,B_2)/B_{1}
\label{eq:HC}
\end{equation}
for small fields $B_1\leq\unit{0.4}\tesla$
(cf. Fig.\ {\ref{fig:RhoHvsB1_new}}). 

The split-coil magnet {generating the tuning field which tunes the material from the magnetic ground state to the paramagnetic one is applied within the magnetic $ab$ plane}. This magnet covers a range of $B_2$  extending to a value as large as 4\,T, nearly two orders of magnitude larger than the $ab$-plane critical field, $B_{2\text c} \approx 0.06$~T. The combination {of the two magnets and the dilution refrigerator} enabled us to access a wide range of both temperature and magnetic field.
This was essential for
extracting the temperature dependence of the FWHM as well as for the separation of the crossover and background contribution{s}.
The fitting of the isotherms was restricted to below \unit{2}\tesla\ for the crossed-field Hall results due to deviations from
linearity at higher fields. These deviations are likely associated with the Zeeman-splitting of the bands, which becomes sizable in this field range {since $B_2$} is applied in the easy magnetic plane.

{In} the case of the single-field Hall
experiment {($B_2=0$)} {(see Fig.~\ref{fig:setup}(b))}, the differential Hall coefficient 
\begin{equation}
	\tilde{R}_{\mathrm H}(B_1) \equiv
\frac{\partial \rho_{\mathrm H}(B_1,0)}{\partial B_1}
\label{eq:TRH}
\end{equation} was  calculated and analyzed.
The fitting of the differential Hall coefficient $\tilde{R}_{\mathrm H}(B_1)$ with eq.~3 of the main text
{leads to} the parameters $\tilde{R}_{\mathrm H}^0$ and $\tilde{R}_{\mathrm H}^{\infty}$. 
In the case of the single-field Hall experiment, the fitting was
performed over the full field range up to \unit{4}\tesla.

Special care was spent to a precise alignment  of the field $B_1$ to be parallel to the $c$ axis within less than 0.5\degree. {Such precise} alignment is essential for the single-field experiment as a small component of this field within the $ab$ plane could easily dominate the tuning due to the large {magnetic} anisotropy of the material.  

The magnetoresistivity $\rho(B_2)$ {(see Fig.~\ref{fig:setup}(c) for a sketch of the setup)} {was monitored during the crossed-field Hall-effect measurements}.
By an analogous fitting with the crossover function the corresponding parameters $\rho^0$ and $\rho^{\infty}$ were extracted. We note that for the magnetoresistivity curves the linear term $mB$ is
negligibly small {(cf.\ section \ref{sec:SampleDep})}.

%
%
\section{Samples}
\label{sec:Samples}
Single crystals of YbRh$_2$Si$_2$ were synthesized using an {indium} flux-growth
technique as described  earlier \cite{Trovarelli_00S}. An additional optimization
of the initial composition and the temperature profile led to an improved sample
quality. This is manifested in an {almost} doubled residual resistance ratio of 
sample 2 {($\mathrm{RRR}=120$)} compared to sample 1 {($\mathrm{RRR}=70$)}.
 All samples were polished to thin
($t\lesssim\unit{80}\micro\metre$)  platelets and prescreened via resistivity
$\rho(T,B)$ measurements to ensure {indium}-free samples.

Two different samples are considered. Sample 1 was taken from
Ref.~\cite{Paschen2004S} and remeasured in the newly designed 
{high-resolution} setup {applying the precise alignment 
procedure described in section \ref{sec:expDetails}} as a cross
check. Sample 2 was chosen from our highest quality batch 
{(see above)}. These two samples span the whole range of sample dependences in the
low-temperature  Hall effect \cite{Friedemann2008S}{:}
the low-temperature Hall coefficient seems to depend on tiny changes of the
composition (samples from the same batch, on the other hand, show identical behavior).

%
%
\section{Anomalous Hall contribution}
\label{sec:AHE}
In heavy-fermion metals, an asymmetric scattering of the conduction electrons from the 4$f$ moments -- the skew scattering -- leads to an anomalous contribution to the Hall coefficient mostly relevant {at high temperatures} \cite{Fert1987S}. However, our analysis of the crossover at the quantum critical point (QCP) is not affected by the anomalous Hall effect, since its contribution to the Hall resistivity 
\begin{equation}
	\rhoH^{\text a}(B_1) = C \rho(B_1)\mu_0 M(B_1)
	\label{eq:AHW}
\end{equation}
is {essentially} linear in field. It adds a small, but constant anomalous contribution $\RH^{\mathrm a}$ to the differential Hall coefficient with an absolute value of less than $\unit{0.07 \times 10^{-10}}\cubicmetre\per\coulomb$. This is negligible compared to the large variation of \drhoHdB\ (compare vertical bars in Fig.~\ref{fig:drhohdb1_YRS_new} and Fig.~\ref{fig:RHvsB2_Silk}(b)). Furthermore, both the inflection point and the sharpness of the crossover are invariant to a constant offset. In eq.~\ref{eq:AHW}, $C$ denotes a constant which was determined from fits to \RT\ {at high temperatures ($\approx\unit{100}\kelvin$ in \YRS)} \cite{Paschen2005} where {no} sample dependences are observed \cite{Friedemann2008S}. The resistivity $\rho(B_1)$ was measured simultaneously, and the magnetization $M(B_1)$ for the relevant geometry ($B \parallel c$) was taken from Ref.~\cite{Gegenwart2002S}. {Since the anomalous contribution does not influence our analysis, we have simply considered the raw data.}

\section{Crossovers in Hall effect and magnetoresistivity}
\begin{figure}
	\centering
		\includegraphics[width=.9\columnwidth]{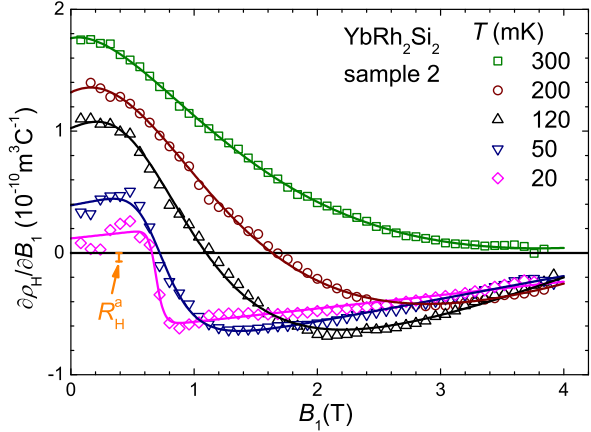}
	\caption{\label{fig:drhohdb1_YRS_new} \textbf{Single-field results 
on sample 2 of \YRS.} Selected isotherms of the numerically derived 
differential Hall coefficient $\TRH(B_1) = {\partial \rhoH}/{\partial B_1}$ for sample 2 are plotted against $B_1$. 
The solid lines are fits of the crossover function ({see methods section of the main text}).
At large fields $B \gg B_0$, the third term of eq.~{3} of the main text 
becomes negligible. Consequently, the Hall resistivity is expected to be 
described by the integral of the first two terms ($\TRHI+m B_1$), \textit{i.e.}\
$\rhoH = c + \TRHI B_1 + {\frac{m}{2}} B_1^2 $ with $c$ denoting the intercept with the
ordinate, $\rhoH(B_1=0)$. This form was consistently
fitted. The {vertical,} orange bar 
represents the anomalous contribution determined from eq.~\ref{eq:AHW} {via the derivative $R_{\mathrm H}^{\mathrm a} = \partial \rho_{\mathrm H}^{\mathrm a} / \partial B_1$} (see {text}).}
\end{figure}

\begin{figure}
	\centering
		\includegraphics[width=.9\columnwidth]{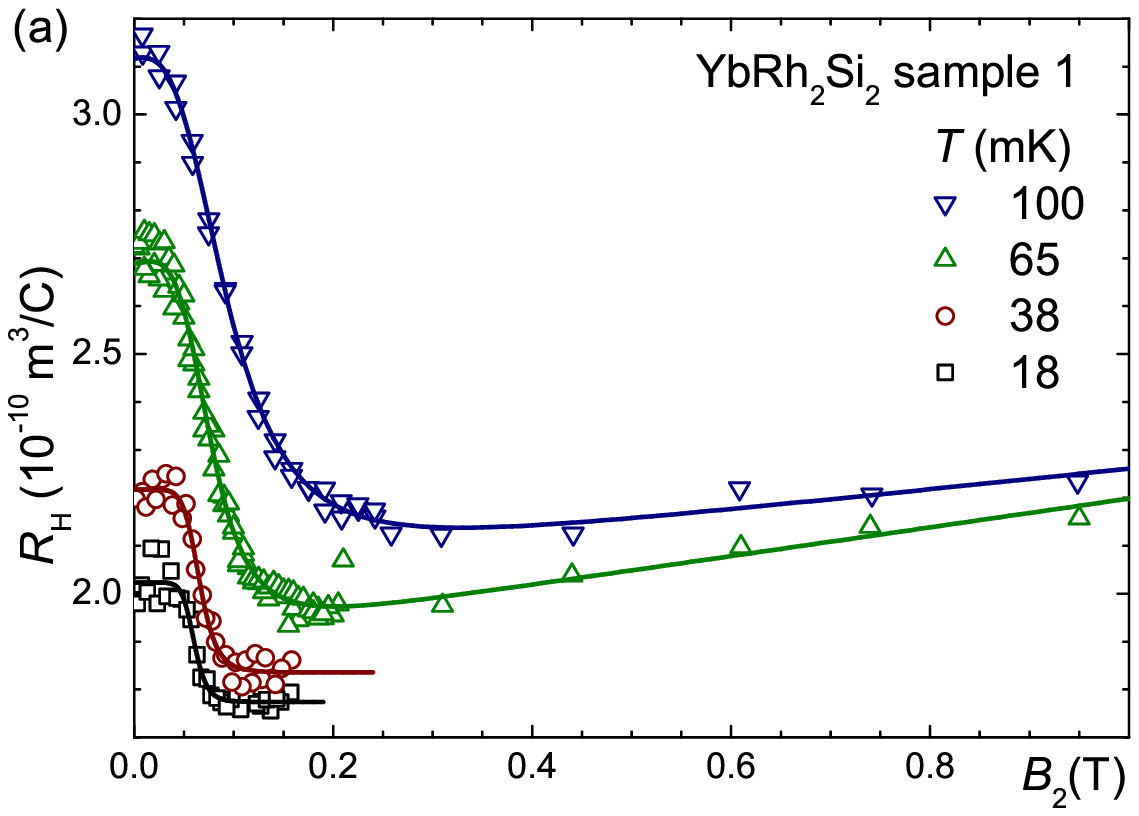}
		\includegraphics[width=.9\columnwidth]{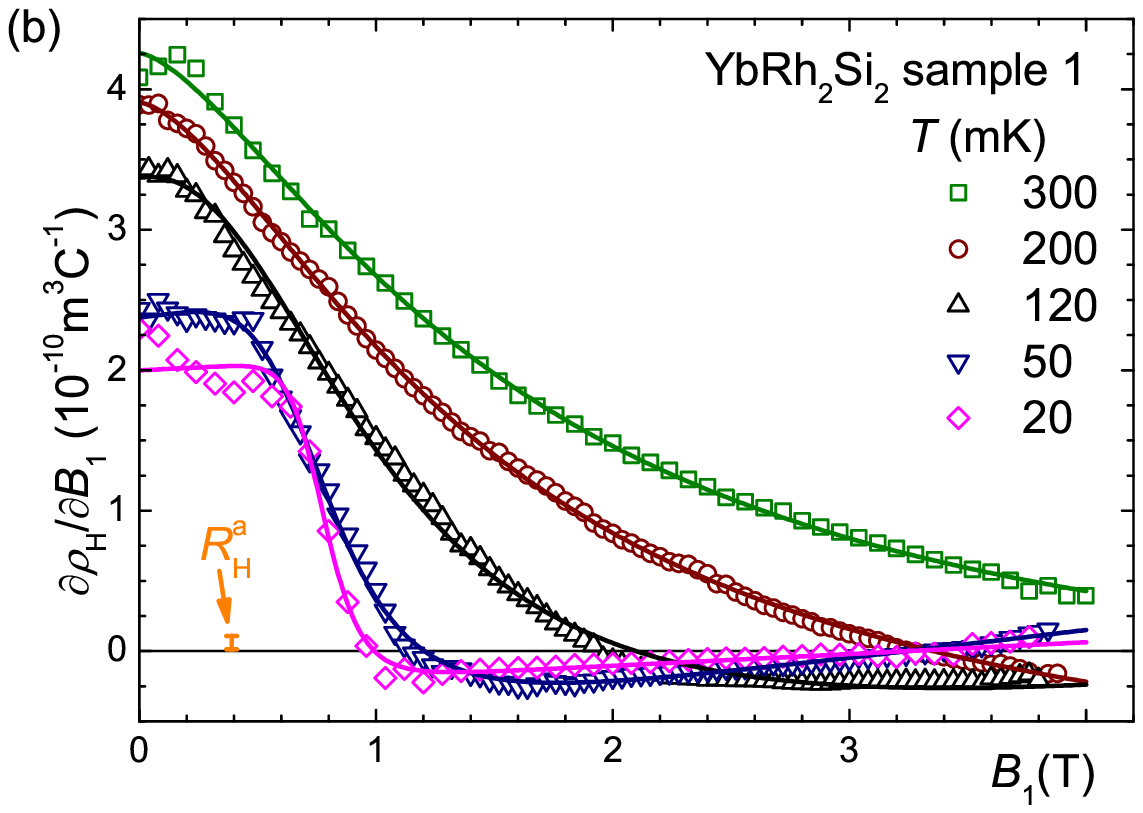}
	\caption{\label{fig:Silkes} {\textbf{Hall effect results on sample 1 of \YRS.}} {\bf (a)} Crossed-field and {\bf (b)} single-field results for the Hall coefficient \RHB\ and the differential Hall coefficient \TRHB, respectively. For a description of the fitting function {we refer} to the Methods section of the main text. As in Fig.\ \ref{fig:drhohdb1_YRS_new}, the anomalous contribution $R_{\mathrm H}^{\mathrm a}$ is indicated by the vertical, orange bar {(see also section \ref{sec:AHE})}.}
	\label{fig:RHvsB2_Silk}
\end{figure}

Figure~\ref{fig:drhohdb1_YRS_new} shows the single-field Hall-effect results of \YRS\ for sample 2. 
The crossover found in the differential Hall coefficient $\TRH(B_1)$ was fitted {by} the same crossover function as in the crossed-field experiment {(eq.~{3} of the main text)}. In particular, it is possible to identify the crossover and  background contribution{s}. The {corresponding} results {and the fits} for the crossed-field ($\RH(B_2)$) and {{the} single-field ($\TRH(B_1)$) experiments} of sample 1 are shown in Fig.~\ref{fig:Silkes}. 

\begin{figure}
	\centering
		\includegraphics[width=.9\columnwidth]{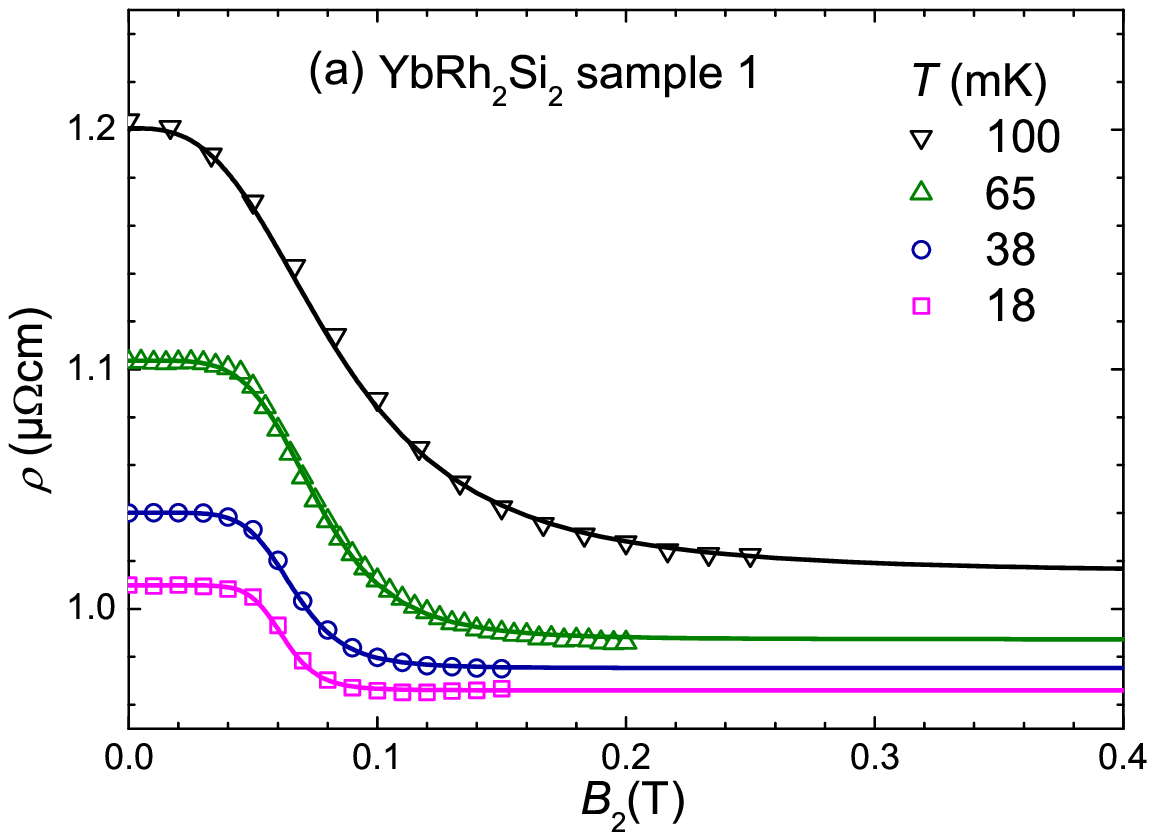}
		\includegraphics[width=.9\columnwidth]{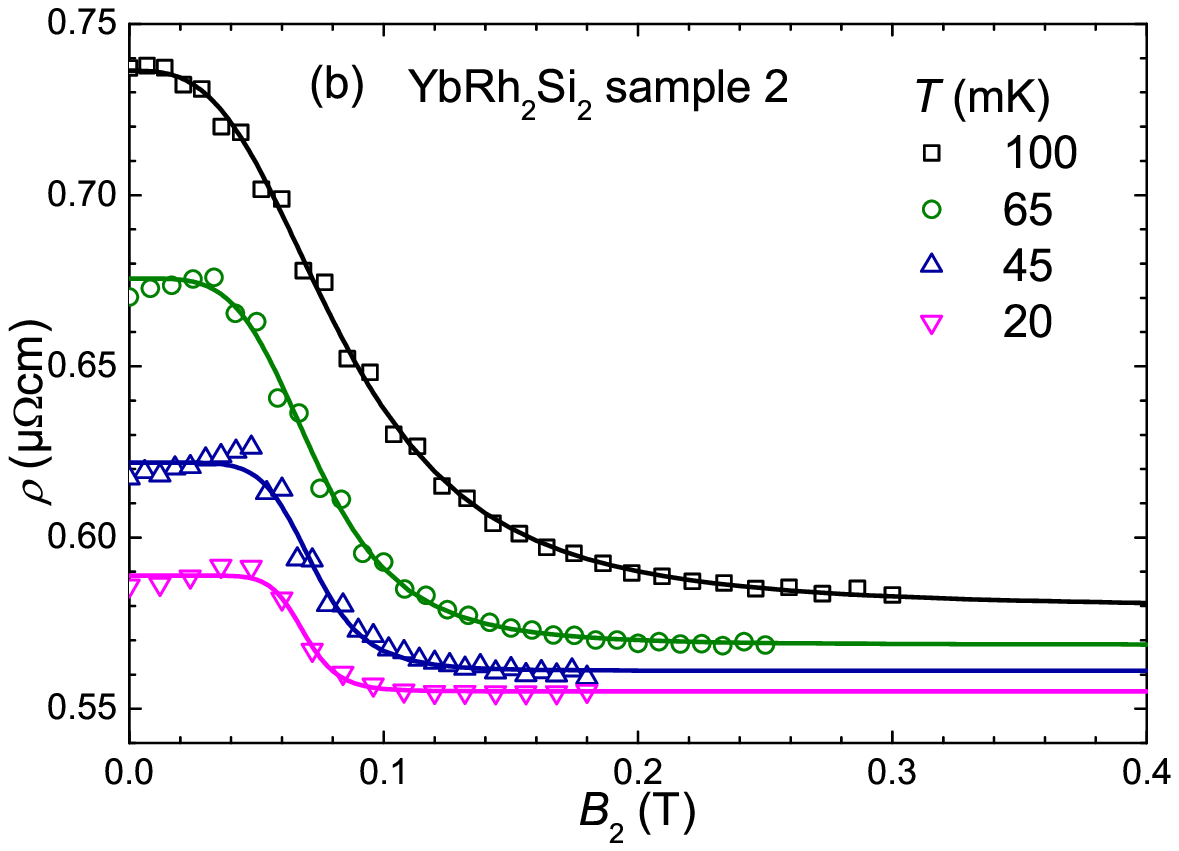}
	\caption{\label{fig:mr_YRS_Silk} {\textbf{Crossover in the magnetoresistivity $\mathbf{\rho(\textit{B}_2)}$ of \YRS.} Results \textbf{(a)} for sample 1 and \textbf{(b)} for sample 2 are depicted.} The data were measured simultaneously {with} the crossed-field Hall effect experiment. For the fitting of the crossover {(eq.\ {3} of the main text)} the linear background term was omitted.}
\end{figure}

\begin{figure}
	\centering
		\includegraphics[width=.9\columnwidth]{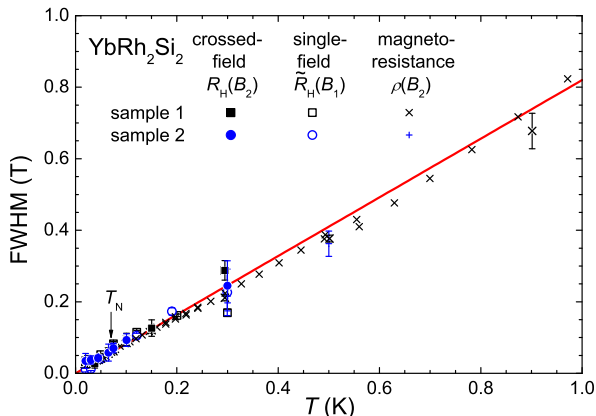}
		\caption{\label{fig:FWHM_Suppl} {{\textbf{ {Full-width} at half-maximum (FWHM) up to 1\,K.} The results extracted from the magnetoresistivity crossover are depicted in an enlarged temperature range up to \unit{1}\kelvin. Solid line represents the very same linear fit as in Fig.\ 3 of the main text. {It is referred} to Fig.\ 3 of the main text for further explanations.}}}
\end{figure}

The magnetoresistivity ($\rho(B_2)$) data are illustrated in Fig.~\ref{fig:mr_YRS_Silk} for sample 1 and 2.
{Here, the same crossover function (eq. 3 of the main text) was used, with the linear term being negligibly small, $m\simeq 0$.} We analyzed the magnetoresistivity crossover up to \unit{1}\kelvin. This allows {us} to extract the FWHM over this enlarged temperature {range} as depicted in Fig.\ \ref{fig:FWHM_Suppl} proving that the width follows the unique linear temperature up to such high temperatures.  

No signature is seen in the temperature dependence of the FWHM for any of the experiments at the N\'eel transition (cf. main text). 
Only below \unit{30}\milli\kelvin\ does the FWHM extracted from both the crossed-field Hall effect and the magnetoresistivity crossovers seem to tend slightly towards larger values compared to the overall linear temperature dependence{, see Fig.\ 3 of the main text}. We assign this to influences arising from the nearby classical phase transition as at this temperature the phase boundary is approached by the Hall crossover. This is seen in the phase diagram with the FWHM around the  crossover field substantially extending into the magnetically ordered phase for temperatures below \unit{30}\milli\kelvin\ only (cf. horizontal bars in Fig.\ 4 of the main text), the temperature below which the seeming deviations from linearity occur. Further indications of this additional influence are observed in the crossed-field Hall coefficient curves which exhibit {visible spread} at lowest temperatures within the ordered phase (Fig.\ 1 of the main text and Fig.\ \ref{fig:RHvsB2_Silk}). 

The FWHM extracted from the single-field measurements on the other hand continues to obey the linear form down to the lowest temperatures accessed. Given this continuity, {we consider the single-field data to} represent the intrinsic quantum critical behavior. This 
property of the single-field experiment may arise from the fact that only here the tuning-field is applied along the magnetic hard axis. For this orientation the magnetization and consequently also the classical magnetic fluctuations are by almost one order of magnitude smaller compared to those for fields applied within the magnetic easy plane \cite{Gegenwart2002S}. The latter configuration is realized in both the crossed-field Hall effect and the magnetoresistivity measurements. 

Finally, we would like to note that within the experimental accuracy the FWHM of all the measurements is compatible with the linear form in the whole temperature range. {Taking all this together, the data imply that the linear temperature dependence of the FWHM represents the behavior intrinsic to the quantum criticality.}

The vanishing FWHM implies a jump of the magnetoresistivity at zero temperature, in contrast to the common behavior of Kondo systems for which the width of the change in magnetoresistivity remains finite at zero temperature \cite{Schlottmann1983}. 
{This represents a key element in our interpretation of the Hall crossover in terms of a Fermi-surface reconstruction.}

%
%
\section{Sample dependence of the background contribution}
\label{sec:SampleDep}
\begin{figure}
	\includegraphics[width=.7\columnwidth]{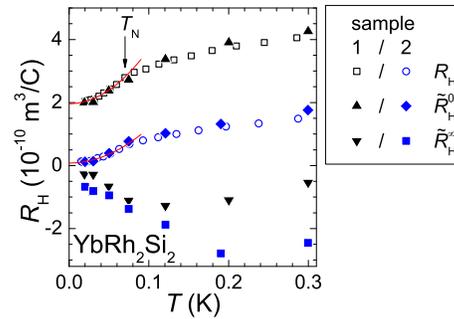}  
	\caption{\label{fig:RHvsTlinS}
\textbf{Limiting values of the single-field Hall crossover.}
{Fit parameters \TRHO\ and \TRHI\ of the single-field Hall-effect measurement $\TRH(B_1)$. {Results for} sample 1 and sample 2 are depicted along with initial-slope Hall coefficient. 
{Solid curves represent quadratic fits as discussed in the text (see also Fig.\ \ref{fig:RHiundRHovsTsq}).} Standard {deviations} are smaller than the symbol size.} 
}
\end{figure}

\begin{figure}
	\includegraphics[width=.9\columnwidth]{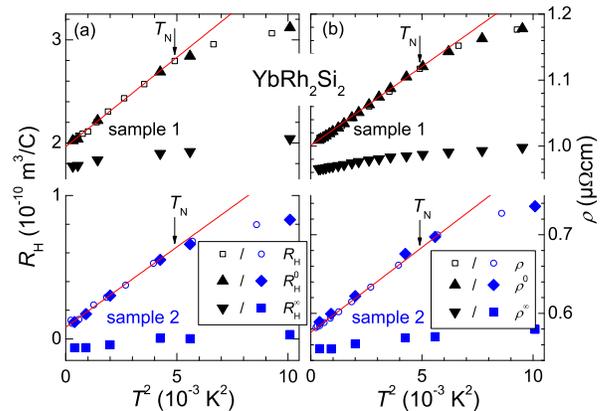} 
	\caption{\label{fig:RHiundRHovsTsq}
{\textbf{Evolution of the Hall coefficient in the antiferromagnetically ordered phase.}
{\textbf{(a)} Initial-slope Hall coefficient \RT\ is plotted against $T^2$ together with the fit parameters \RHO\ and \RHI. Solid lines represent fits of the form $\RT = c+A^{\prime}\usk T^2$ for temperatures below \TN\ (indicated by arrows) {where $c$ denotes the intercept with the ordinate, i.e. the zero-temperature Hall coefficient, $\RH(T=0)$}. These {fits} are reproduced as solid curves in Fig.~2(a) of the main text and in Fig.~\ref{fig:RHvsTlinS}. \textbf{(b)} Corresponding plot of the resistivity $\rho(T)$ and the fit parameters \rhoO\ and \rhoI\ extracted from the magnetoresistivity crossover with an according fitting function $\rho(T)=\rho_0 + A \usk T^2$ {($\rho_0$ being the residual resistivity)} below \TN\ {which agrees well with previous observations \cite{Gegenwart2002S}}. } 
}}
\end{figure}

The temperature dependences of \RHO\ and \RHI\ are shown in Fig.\ {2(a) of the main text} for both samples.
The results of the low-field value \RHO\ are in good agreement with those of the {measured} zero-field ($B_2=0$) initial-slope Hall coefficient \RT\ (Ref.~\cite{Friedemann2008S}). This is non-trivial as \RHO\ is the result of the fitting procedure {specified in the Methods section of the main text}, and not fixed to the zero-field value. Both, $\RHO(T)$ and $\RHI(T)$ decrease as the temperature decreases. 
{Below $T_{\mathrm N}$, the Hall coefficient clearly obeys a quadratic temperature dependence as demonstrated in Fig.\ \ref{fig:RHiundRHovsTsq}(a). This finding allows for a proper extrapolation of \RT\ and $\RHO(T)$ to $T \to 0$ which in turn enables us to conclude that the difference between \RHO\ and \RHI, \textit{i.e.}\ the height of the Hall crossover, remains finite for both samples on approaching zero temperature.}
Identical observations were made for the parameters extracted from the magnetoresistivity crossover (Fig.\ \ref{fig:RHiundRHovsTsq}(b) and Fig.\ 2(b) of the main text) as well as for those of the crossover in {the} single-field Hall effect (Fig.\ \ref{fig:RHvsTlinS}). 

\begin{figure}
	\centering
		\includegraphics[width=.8\columnwidth]{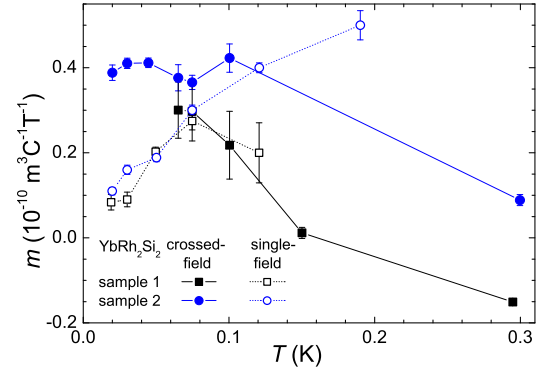}
	\caption{\label{fig:MvsT} Temperature dependence of $m$, 
	the slope of the linear background contribution {extracted from the crossed-field and single-field Hall-effect measurements.}}
\end{figure}

The {temperature} evolution of the parameter $m$ 
{of eq.~{3} of the main text}
{(which describes the linear background)} is shown 
in Fig.~\ref{fig:MvsT} for all Hall-effect experiments. The data for the 
single-field experiments are restricted to below 0.2\,K due to the
crossover extending over an increasingly large field range with increasing
temperature. This prevents a proper determination of the background contribution 
within the field range of our experiment (cf. Fig.~\ref{fig:drhohdb1_YRS_new}
and Fig.~\ref{fig:Silkes}(b)). The crossed-field results of sample 1
are limited to temperatures above \unit{0.065}\kelvin\ due to 
a lack of high-field data at lower $T$ (cf. Fig.~\ref{fig:Silkes}(a)).

Taking all these together we find a pronounced sample dependence for the quantities describing the background contribution whereas those associated with the critical crossover are {essentially sample independent} (cf.\ main text).

%
%
\section{Single-electron Green's function and the Hall crossover}
\label{sec:scaling-vs-Hall}
The single-electron Green's function, $G({\bf k},E,T)$ (eq.~1 of the main text) 
on either side of the zero-temperature transition and at low temperatures,
can be decomposed as
\begin{equation}
G({\bf k},E,T)=G_{\mbox{\small coh}}({\bf k},E,T)+G_{\mbox{\small inc}}({\bf k},E,T).
\label{eq:GreensFunction}
\end{equation}
This decomposition is an immediate consequence of the fact that the phases
separated by the QCP are taken to be Fermi liquids. Indeed, because of 
the jump of the Hall coefficient across the QCP, the Fermi liquids are 
taken to have large and small Fermi surfaces, respectively.
The coherent part of $G({\bf k},E,T)$ is given by
\begin{equation}
G_{\mbox{\small coh}}({\bf k},E,T)=\frac{z_{\bf k}}{E-\varepsilon({\bf k})+i\Gamma_{\bf k}(T)}
\label{eq:GreensCoherent}
\end{equation}
describing a quasiparticle,
and $G_{\mbox{\small inc}}({\bf k},E,T)$ is a background contribution.
The strength of the quasiparticle excitation,
$z_{\bf k}$, formally defined as the residue
of the pole,
is non-zero in either phase.
However, to be compatible with the continuous nature of the zero-temperature
transition, $z_{\bf k}$ must 
vanish as the QCP is approached: $z_{\bf k}
\rightarrow 0$ as $B\rightarrow B_c$.
At $T=0$, the quasiparticle damping
$\Gamma_{\bf k}$ vanishes at the small Fermi-momenta ${\bf{k}}_{\mathrm F}$ for $B<B_c$,
and at the large Fermi-momenta ${\bf{k}}_{\mathrm F}^*$ for
$B>B_c$; at these respective Fermi-momenta, the quasiparticles become 
infinitely-sharp
excitations at zero temperature.
The coherent part of $G({\bf k},E,T)$ is therefore the 
diagnostic feature on either side of the transition, and it jumps at the QCP
in accord with the sudden Fermi surface change.

This jump is manifested in the Hall measurement, because 
the Hall coefficient is independent of the quasiparticle residue.
{Note that our argument 
builds on the 
Landau 
Fermi-liquid nature of the phases on
either side of the
QCP, where the Fermi surface and the forms (\ref{eq:GreensFunction})
and (\ref{eq:GreensCoherent}) are well defined.
The Hall coefficient of a Fermi liquid is completely determined
by the dispersion of the single-electron excitations near the Fermi surface
\cite{Kohno1988,Khodas_03S}.
The fact that the quasiparticle residue does not appear in the Hall coefficient
of a Fermi liquid can be seen in a number of related ways.
It is known -- both phenomenologically \cite{Pines} and
microscopically \cite{Nozieres} -- that the Boltzmann equation 
of a Fermi liquid does not depend on the quasiparticle residue; by extension,
the Hall coefficient does not depend on the quasiparticle residue.
The same conclusion
is reached through a study of the Hall coefficient of a Fermi liquid using 
the Kubo formula \cite{Kohno1988}.
Finally, for
a spherically symmetric but otherwise arbitrary dispersion 
it has been 
shown explicitly by
diagrammatic
calculations that the Hall coefficient is not renormalized by the 
electron-electron interactions}
\cite{Khodas_03S}.

At non-zero temperatures
, the quasiparticle relaxation rate at either
${\bf{k}}_{\mathrm F}$ for $B<B_c$, 
or ${\bf{k}}_{\mathrm F}^*$ for $B>B_c$, 
no longer vanishes.
In fact, inside the Fermi-liquid
phase (with either large or small Fermi surface), the temperature dependence
of  $\Gamma_{\bf k_{F}}$ has to be quadratic in $T$.
{However, the Fermi surface remains well defined in these regimes. 
At finite temperatures,
the change from one Fermi surface to the other is therefore 
restricted to the intermediate quantum critical regime. 
Because of the absence of a phase transition at any non-zero temperature,
the sharp reconstruction of the Fermi surface at $T=0$ is turned into 
a Fermi-surface crossover across the $T^*(B)$ line.
The restriction to the quantum critical regime implicates that the linear in temperature
relaxation rate present in this regime determines the broadening of the Fermi
surface change.
From the relation of the Hall coefficient to the Fermi surface
(described above) we associate the width of the Hall crossover with this 
broadening and consequently with the relaxation rate $\Gamma$ of the 
single-electron Green's function.}

To be more specific, consider
a general scaling form for the single-electron Green's 
function at the Fermi momentum in the quantum critical region:
\begin{equation}
G(E,T) =\frac{1}{T^\alpha} \phi \left(\frac{E}{T^x} \right).
\end{equation}
{For a Gaussian fixed point, a dangerously irrelevant variable 
will invalidate $E/T$-scaling and make $x>1$. Correspondingly, 
the single-particle relaxation rate $\Gamma$ (defined in eq.~4 
of the main text) 
will be superlinear in temperature and the Hall crossover width as 
a function of the magnetic field will in general not be linear 
in temperature. An} {interacting}
{fixed point, on the other hand, can generate $x=1$ with a relaxation rate that is linear in temperature and}
{be} {compatible with the linear-in-$T$ Hall-crossover width observed.}

{An important question concerns the critical exponent $y$ as defined
by the $T^*(B)$ line, $T^*(B) \propto (B-B_c)^y$. This scale
equivalently specifies $B^*(T)$, the center of the critical Hall 
crossover. In general, $B^*$ and FWHM are two independent parameters. 
Eq.~5, which fits our data very well (Figs.~1, S3, S4, S5), invokes
both parameters to characterize the critical component of the Hall
crossover.} 
{We have already shown that the FWHM is robustly linear in temperature.
For the $T^*(B)$ line, Fig.~4 of the main text suggests that 
the exponent $y$ is less than $1$ but its precise determination,
especially from the Hall-effect measurements, requires accuracies
beyond our present experiments.}



\end{article}

\end{document}